\documentclass[reprint, aps, prx]{revtex4-2} 

\usepackage{amsmath}
\usepackage{amssymb}
\usepackage{bm}        
\usepackage{graphicx}  
\usepackage{dcolumn}   
\usepackage{booktabs}  
\usepackage{array}
\usepackage{xcolor}
\usepackage{hyperref}
\hypersetup{
    colorlinks=true,
    linkcolor=red,     
    citecolor=blue,    
    urlcolor=cyan      
}

\begin{document}
\title{Experimental demonstration of the Quantum Fourier Transform on up to 100 qubits using a convolutional compilation strategy}

\author{Paul Coote}
\email{paul.coote@q-ctrl.com}
\author{Michael J. Biercuk}
\author{Yuval Baum}
\affiliation{Q-CTRL, Los Angeles, CA USA and Sydney, NSW Australia}
\date{September 28, 2026}

\begin{abstract}
We present and experimentally validate the `Convolutional QFT': a constructive compilation strategy for the Quantum Fourier Transform (QFT) subroutine on a linear nearest neighbor (LNN) qubit topology. We first introduce a novel strategy that compiles the $n$-qubit QFT onto an LNN topology using only $n^2 - n$ $CX$ gates, matching requirements of a direct compilation on an all-to-all architecture. We then derive the convolutional variant used in our experiments, which requires an additional two $CX$ gates in total, and is realized via a compact, translation-invariant kernel circuit gadget that traverses a quantum register. We demonstrate the power of the convolutional compilation strategy on the IBM Quantum Platform by executing QFT benchmarking circuits. We measure a process fidelity of 11.4\% at 50 qubits, and 1.8\% at 80 qubits. The correct output state remains clearly distinguishable above background noise up to 100 qubits. These results constitute the largest experimental QFT demonstrated on any quantum computing hardware to date.
\end{abstract}

\maketitle

\section{Introduction}

Compiling quantum algorithms for devices with limited qubit connectivity introduces substantial resource overhead \citep{preskill2018quantum,peham2023optimal,herbert2020depth}. Operations defined in abstract algorithms must be remapped to the specific native gate set and restricted interaction topology of the target device \cite{sinha2022qubit, hua2023qasmtrans}. In many hardware architectures, direct entangling operations are restricted to proximate qubits; for example, neighboring physical qubits \cite{arute2019quantum,kim2023evidence}, adjacent error-correcting code patches \citep{fowler2012surface,litinski2019game}, or qubits that are moved together using shuttling \citep{bluvstein2022quantum,kielpinski2002architecture}. Executing long-distance interactions requires routing quantum information across the chip, which introduces additional entangling gates \cite{wille2023mqt,li2019tackling}. These extra operations degrade performance in near-term devices \cite{gokhale2024faster,karuppasamy2025comprehensive}, and inflate the spacetime volume in fault-tolerant architectures \cite{hamada2026efficient,herzog2025exploiting}.

Consequently, execution of quantum algorithms on physical hardware benefits from highly optimized, device-aware compilation. Combining optimized compilation with error suppression techniques such as dynamical decoupling dramatically improves algorithmic performance \cite{coote2025resource,mundada2023experimental}. State-of-the-art demonstrations of quantum algorithms utilize both strategies in order to deliver performance at the threshold of quantum advantage on pre-fault-tolerant devices \cite{hartnett2026fast,liao2025achieving,sachdeva2026integrated}.  Going beyond application-agnostic approaches, it becomes attractive to explore targeted compilation strategies for specific subroutines relevant to large-scale quantum algorithms, such as the quantum Fourier transform (QFT)~\citep{nielsen2010quantum,nam2020approximate}, with the objective of reducing resource requirements at scale~\citep{fowler2004implementation, park2023reducing,park2025reducing,dreier2025connectivity}.

In this work, we introduce a novel compilation strategy for the QFT subroutine that dramatically reduces resource requirements and improves performance in hardware execution.  We first derive a compilation of the $n$-qubit QFT based on the low-connectivity linear nearest neighbor (LNN) architecture that requires $n^2-n$ $CX$ gates. This matches the $CX$-gate-scaling achieved for a direct compilation for an all-to-all connectivity architecture~\cite{holmes2020impact}, and persists for any truncation threshold of the approximate QFT (AQFT) \citep{fowler2004scalability,nam2013scaling,nam2013streamlining,coppersmith2002approximate,white2023truncated}, effectively eliminating all routing overhead. We further identify a refinement to improve experimental fidelity: by adding a single ancilla qubit and two additional $CX$ gates, the circuit can be transformed into the form that we call the `Convolutional QFT' or `Convolutional AQFT'. In this construction, the algorithm can be synthesized using a compact circuit kernel gadget, which incrementally steps along the quantum register and reduces the average number of two-qubit gates in the causal history (or `light cone' \cite{cai2023quantum,eddins2024lightcone}) of each qubit.  

Experimental demonstrations on the IBM Quantum Platform \cite{IBMQuantumPlatform} reliably identify the correct frequency of a Hilbert space signal for circuits with width up to 100 qubits. We achieve a process fidelity above 10\% at a width of 50 qubits, and above 1\% for 80 qubits. In all test circuits, the correct 50-bit or 80-bit integer frequency of the periodic quantum state is extracted with high selectivity, specificity, and reproducibility. Beyond an 80-qubit circuit width, the process fidelity drops below the conventional 1\% threshold used in the literature~\cite{baumer2024quantum,aumann2026demonstrating}, but up to $n=100$, the target bitstring remains the most frequently observed state compared to any incorrect candidate.

\section{Background}

\begin{figure*}
    \centering
    \includegraphics[width=0.99\linewidth]{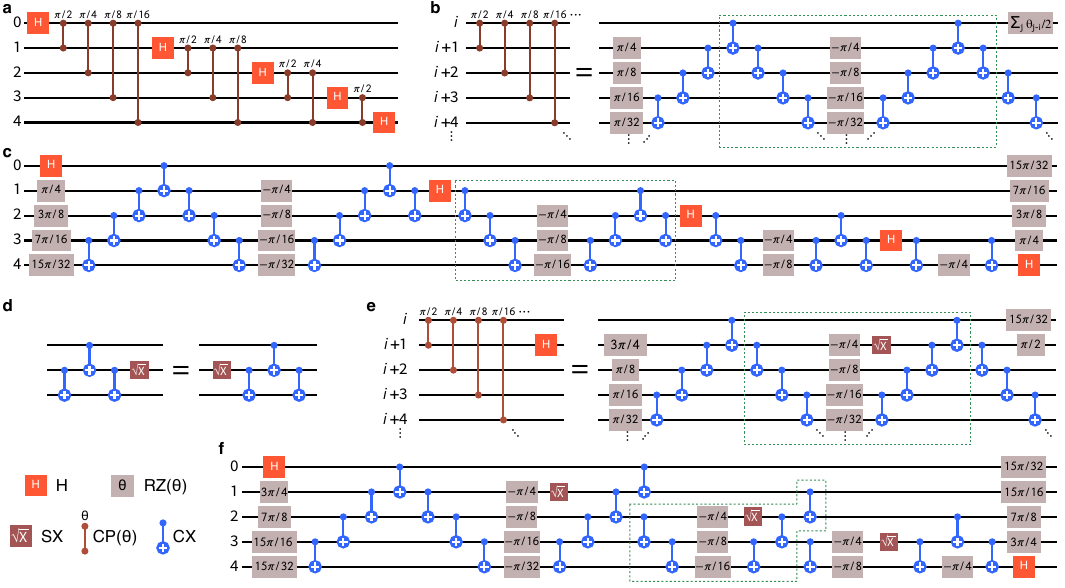}
    \caption{Derivation of the Low-CX QFT on an LNN topology. {\bf a.} The textbook QFT algorithm for $n=5$ qubits comprising layers of Hadamard and $CP$ gates. {\bf b.} A realization of a layer of $CP$ gates using $CX$ and $RZ$ gates compiled to an LNN topology, adapted from \cite{park2023reducing}. 
    Only the $CX$ within the green dashed box are required for the QFT, since other $CX$ cancel with adjacent layers. 
    {\bf c.} The Park/Ahn QFT for 5 qubits, requiring 26 $CX$ gates. 
    Each layer of $d$ rotations requires $2d+1$ $CX$ gates (green dashed box).
    {\bf d.} The $\sqrt{X}$ gate commutes with a configuration of three $CX$ gates. 
    {\bf e.} A realization of the $CP$ gates for layer $i$ of the QFT, followed by the Hadamard gate of the next ($i+1$) layer. Only the gates within the green dashed box are required for the QFT. 
    {\bf f.} Low-$CX$ LNN compilation of the QFT for 5 qubits, requiring 20 $CX$ gates (six fewer than panel c). Each layer of $d$ rotations requires $2d-1$ $CX$ gates (green dashed region).}
    \label{fig:lowCX}
\end{figure*}

The unitary of the n-qubit QFT \cite{nielsen2010quantum} is given by
\begin{align}
    \text{U}_{\text{QFT}} (n) = \prod_{i=1}^{n} H_i \prod_{j=i+1}^{n} CP_{i,j}(\theta_{j-i}) \label{eq:qft_unitary}
\end{align}
where $i$ and $j$ index the qubit wires, $H_i$ is a Hadamard gate, $CP_{i,j}$ is a controlled-phase gate, and the phase angles are $\theta_{k} = \pi / 2^{k}$. These angles form an exponential hierarchy, with $n-k$ angles of size $\theta_k$, for $k \in [1, n)$. A circuit diagram for the case $n=5$ is given in Fig \ref{fig:lowCX}a. The AQFT omits $CP$ gates with angles smaller than a given approximation threshold. 

On an idealized all-to-all connected architecture, each controlled-phase gate $\text{CP}(\theta_k)$ can be synthesized using two $\text{CX}$ gates and three single-qubit $R_Z$ gates \citep{barenco1995elementary, nielsen2010quantum}. The full QFT including all angles requires $n^2-n$ $CX$ gates using a direct synthesis of each $CP$ gate. This $CX$ complexity is reported frequently in the literature \citep{holmes2020impact,klaver2026swap, aumann2026demonstrating}; recently, a more efficient synthesis of the full QFT has been suggested for an all-to-all connected device using a parity-encoding technique\cite{dreier2025connectivity}. 

Because all connected graphs can embed a linear subgraph,  we  focus on LNN compilation strategies as generic tools that can be used to compile any quantum algorithm to any connected quantum processor.  Our approach builds upon an earlier compilation scheme proposed by Park and Ahn \cite{park2023reducing, park2025reducing}, which we briefly review for context.

Park and Ahn describe a compilation strategy that realizes each layer of $CP$ gates in the QFT on an LNN topology \cite{park2023reducing}, using four linear chains of $CX$ gates and a column of $RZ$ in the center (Fig \ref{fig:lowCX}b). The motif also includes $RZ$ gates at the start (end), which can be moved to the start (end) of the entire QFT circuit, since they commute with preceding layers of $CP$ gates. At the interface of adjoining QFT layers, many $CX$ gates are logically canceled by use of the identity $CX^2=I$. This is possible because $CX$ gates required for one layer are exactly aligned and adjacent to other $CX$ that are part of the preceding or subsequent layer of $CP$ gates (Eq \eqref{eq:qft_unitary}).

In the AQFT, the $d$ largest angles of the QFT are included while others are removed from the circuit. Park and Ahn's construction is directly compatible with the AQFT, and substantial two-qubit-gate savings are directly accessible by omitting negligible rotations. Each removed $CP$ gate from the logical definition of the QFT corresponds to removing an $RZ$ between two adjacent and aligned $CX$. Without the separator, the pair simplifies to the identity and can be removed. The quadratic complexity of the full QFT is thus replaced by a linear scaling, which has slope $2d+2$ (Table \ref{tab:compilation_strategies}). The linear scaling is especially beneficial in the regime $n \gg d$. The resulting circuit uses just $2n-4$ more $CX$ gates for a LNN device than for an ideal all-to-all connected device, for any approximation threshold of the AQFT, up to and including the full QFT (Table \ref{tab:compilation_strategies}). As device performance improves, approximation of the QFT subroutine is anticipated to remain standard even in the fault tolerant era \citep{fowler2004scalability,nam2013scaling,nam2013streamlining,coppersmith2002approximate,white2023truncated}. 

In the next section, we present an alternate LNN compilation exhibiting further reductions of exactly $2n-4$ $CX$ gates, matching the QFT/AQFT $CX$ complexity directly to the all-to-all architecture case.  That is, the novel strategy we introduce completely removes routing overhead, delivering maximum efficiency for arbitrary device architectures.  

\section{Derivation of the Convolutional QFT Circuit}
\subsection{LNN QFT with minimal $CX$ gates}

In this subsection, we leverage circuit identities to find a further reduction in $CX$ complexity. We apply the following three circuit manipulations:
\begin{enumerate}
    \item Each $H$ in the QFT is expressed using the equivalent (up to global phase) synthesis $S\cdot\sqrt{X}\cdot S$, where $S$ is the phase gate.
    \item The two introduced phase gates on each wire are moved all the way to the front or end of the circuit, respectively. This is permitted since $S$ has a diagonal unitary, and commutes with other diagonal sub-circuits including the layers of $CP$ gates present in the QFT.
    \item Each $\sqrt{X}$ gate is moved to the left through three preceding $CX$ gates.
\end{enumerate}

The third step is underpinned by the circuit identity allowing a $\sqrt{X}$ gate to commute through a particular arrangement of three $CX$ gates (Fig. \ref{fig:lowCX}d). The resulting circuit compilation for a single layer of the QFT is depicted in Fig \ref{fig:lowCX}e. For each layer within the QFT, two $CX$ that were previously separated by an $H$ gate are instead adjacent and aligned, and can be removed. The full QFT for $n=5$ is given in Fig \ref{fig:lowCX}f, saving  two $CX$ gates at each of $n-2$ interfaces between adjacent layers, for an overall saving of $2n-4$ $CX$ gates.

\begin{figure*}
    \centering
    \includegraphics[width=0.99\linewidth]{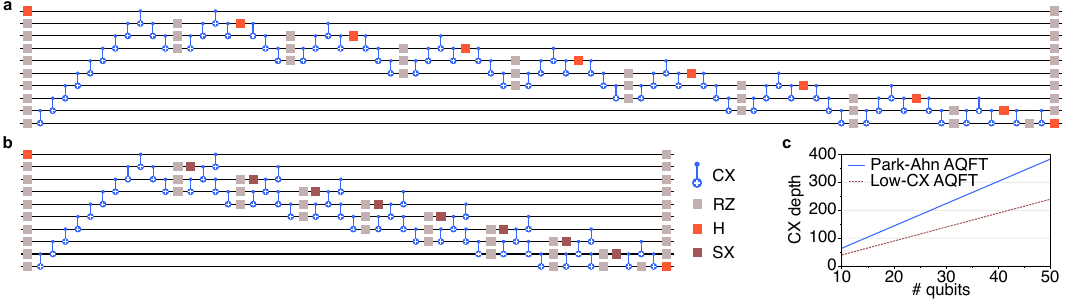}
    \caption{The AQFT on an LNN topology. Single-qubit gate labels and parameter angles are omitted for visual clarity. {\bf a. } The Park and Ahn AQFT layout for $n = 10$ qubits using an approximation threshold of $d = 3$, requiring 64 $\text{CX}$ gates.
    {\bf b.} Our simplified low-CX AQFT with only 48 $CX$ gates. The circuit $CX$ depth is further reduced by additional gate parallelization. {\bf c.} CX-depth of the two circuit compilations, including the $d=3$ largest angles in the QFT. The slope is reduced from $8$ to $5$ by a combination of gate cancellation and parallelization.  }
    \label{fig:simplified}
\end{figure*}

By eliminating these $2n - 4$ gates from the Park/Ahn construction, our LNN compilation strategy achieves exact $CX$ parity with the standard all-to-all implementation, where each $\text{CP}$ gate is synthesized directly using two $\text{CX}$ gates (Table \ref{tab:compilation_strategies}). Remarkably, this construction matches the $\text{CX}$ count of a fully connected topology while operating strictly on an LNN architecture. As a concrete example, Fig \ref{fig:simplified} illustrates an $n = 10$ qubit AQFT that retains the $d = 3$ largest rotation angles in Eq \eqref{eq:qft_unitary}. While the Park and Ahn construction (Fig \ref{fig:simplified}a) requires 64 $CX$ gates, our optimized compilation (Fig \ref{fig:simplified}b) achieves identical algorithmic logic using only 48 $CX$ gates---a reduction of 16 $CX$ gates.

In addition to reducing the $CX$ count, the layout in Fig \ref{fig:simplified}b improves circuit parallelization compared to Fig \ref{fig:simplified}a. Logical cancellation eliminates two gates per qubit, while a third gate can now be executed in parallel with its predecessor, rather than sequentially. Consequently, the linear growth of the circuit depth is reduced by $3n$ (Fig \ref{fig:simplified}c).

Compared to other recent LNN constructions, our synthesis reduces the linear complexity term, but maintains the same leading quadratic order. Specifically, where Park and Ahn \cite{park2023reducing} require $n^2 + n - 4$ $CX$ gates and Klaver et al. \cite{klaver2026swap} achieve $n^2 - 1$, our layout reaches $n^2 - n$ $CX$ gates and matches the standard baseline for all-to-all connectivity (Table \ref{tab:compilation_strategies}). This linear reduction is particularly significant for the AQFT, in which complexity scales linearly for a fixed approximation threshold $d$. For any choice of $d$, our scheme matches the all-to-all baseline, and saves $2n-4$ $CX$ gates over Park and Ahn’s construction. In contrast, Klaver et al.’s parity twine LNN compilation retains quadratic scaling even for the AQFT \cite{klaver2026swap,aumann2026demonstrating,dreier2025connectivity}. This optimization makes our approach exceptionally relevant for both NISQ-era and fault-tolerant hardware, where AQFT variants with a surprisingly high level of approximation enable key algorithms like Shor's to run within practical runtime budgets while retaining high output fidelity \cite{fowler2004scalability,nam2013scaling,nam2013streamlining,coppersmith2002approximate,white2023truncated}. 

Conventional AQFT implementations discard $\text{CP}(\theta)$ gates below a given angular threshold \cite{coppersmith2002approximate}. However, synthesis fidelity can be enhanced at zero $\text{CX}$ overhead by exploiting the decomposition of $\text{CP}(\theta)$ into $R_{ZZ}(-\theta/2) \cdot R_{ZI}(\theta/2) \cdot R_{IZ}(\theta/2)$ (up to a global phase) \cite{nielsen2010quantum}. The two single-qubit rotations are absorbed into $RZ$ gates at the boundaries of our LNN QFT circuits (Table \ref{tab:compilation_strategies}), leaving only the $R_{ZZ}$ term to be truncated. Preserving single-qubit phase contributions yields a closer approximation to the target $\text{CP}$ gate than full gate removal \cite{white2023truncated}.

\begin{table*}
  \centering
\setlength{\tabcolsep}{10pt}
\begin{tabular}{lcccc}
    \toprule
    \textbf{QFT Compilation } & \textbf{CX count} & \textbf{CX count} & \textbf{CX depth}& \textbf{CX depth} \\
    \textbf{Strategy}&\textbf{QFT}&\textbf{AQFT}&\textbf{QFT}&\textbf{AQFT} \\
    \midrule
    All-to-all \cite{holmes2020impact}         & $n^2-n$     & $d(2n-d-1)$   &$4n-6$& $4n-6$   \\
    Park/Ahn \cite{park2023reducing} & $n^2+n-4$       & $\,\,d(2n-d-1)+2n-4$   &$n^2+n-4$& $\,\,2(d+1)n -d(d+1)-4$          \\
    Klaver et al \cite{klaver2026swap}    & $n^2-1$            & $n^2-1$      &$4n-4$& $4n-4$   \\
    Low-$CX$ LNN      & $n^2-n$         & $d(2n-d-1)$        &$n^2-2n+3$&  $(2d-1)n-d(d+1)+3$  \\
    Convolutional        & $n^2-n+2$       & $d(2n-d-1)+2$  &$n^2-3n-8$&  $2(d-1)n-d(d+1)+8$ \\
    \bottomrule
  \end{tabular}
\caption{{Comparison of $CX$ requirements for a selection of QFT compilation strategies. The AQFT includes the $d$ largest controlled-phase angles. Apart from the top row, all strategies are for an LNN qubit connectivity. The last two rows are novel constructions described in this work, and achieve $CX$ gate counts comparable with the all-to-all baseline.}}
  \label{tab:compilation_strategies}  
\end{table*}

\subsection{The Convolutional AQFT} \label{subsec:convolutional_qft}

In this subsection, we perform further circuit transformations to derive the `Convolutional' AQFT. In this circuit, the required entangling operations for the AQFT are realized as a compact circuit gadget. The gadget is a convolutional kernel: it is translation-invariant and steps along the quantum register with a stride length of 1. 

The convolutional strategy is motivated by a key bottleneck in Fig \ref{fig:simplified}b: the initial chain of $n-1$ $CX$ gates. Each gate in this initial chain is effectively upstream of {\em all} other $CX$ gates in the circuit. Execution of this chain on real hardware introduces a level of noise and error, which forms part of the causal history of all downstream logic. This noise level ultimately degrades the overall experimental fidelity. Here, we show that use of an ancilla allows the chain to be reversed, and facilitates a more compact compilation.

\begin{figure*}
    \centering
    \includegraphics[width=0.99\linewidth]{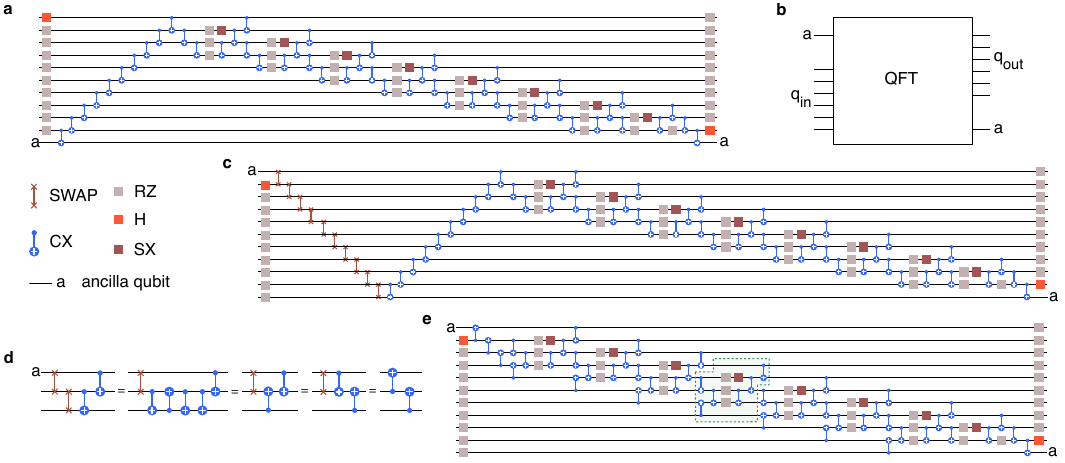}
    \caption{Derivation of the Convolutional AQFT. Single-qubit gate labels and angles are omitted for visual clarity. 
    {\bf a. }  An AQFT for $n=10$ qubits ($d=3$ largest angles) based on Fig \ref{fig:simplified}b, with an additional ancilla qubit wire $a$. {\bf b.} Schematic of a linear-chain AQFT with an ancilla. The data qubits jointly undergo a bit shift of $-1$ along the linear chain during the QFT algorithm. 
    {\bf c.} Insertion of an initial SWAP sequence, which propagates the ancilla across the register. {\bf d.} A chain of SWAPs and an ancilla in the $|0>$ state reverses the sense of a linear chain of $CX$, as explained in the main text. {\bf e.} The Convolutional AQFT. The circuit comprises a compact translation-invariant kernel gadget (green dashed box) that spans $d+2$ adjacent qubits and increments along the register with a stride length of 1. }
    \label{fig:conv}
\end{figure*}

The derivation is summarized in Fig \ref{fig:conv}. We begin by introducing an ancilla initialized in the $\vert{}0\rangle$ state, which lengthens the required LNN chain by one qubit, as in Fig \ref{fig:conv}a. The ancilla is added at the $i=n$ terminal end of the register. The ancilla is targeted by two additional $CX$, which do not change the underlying algorithmic logic. During the algorithm, the ancilla will traverse the register (Fig \ref{fig:conv}b). Next, we introduce a chain of SWAP gates near the start of the circuit to effect this ancilla traversal, shown in Fig \ref{fig:conv}c. In this circuit, the ancilla begins at the top of the register, and is returned at the bottom. The data qubits are each shifted upwards one position by the SWAP gates. 

The additional SWAP gates increase the apparent depth and gate resources of the circuit in Fig \ref{fig:conv}c. However, in this setting the composition of a downward chain of SWAP with an upward chain of $CX$ allows significant simplification. SWAP gates can be synthesized using three adjacent $CX$ with alternating directions. This generates a cancellation of two aligned and adjacent $CX$, as depicted in \ref{fig:conv}d. Another gate cancellation arises from the ancilla. On each wire, the first $CX$ in the SWAP is controlled by the ancilla qubit in the state $|0\rangle$, and is therefore not activated and can be removed. The gates that remain form a linear chain with opposite direction compared to Fig \ref{fig:conv}a. We emphasize that the ancilla is transmitted along the chain of SWAPs; therefore, the leftmost $CX$ in each SWAP is not activated and can be removed from {\em all} wires. 

In Fig \ref{fig:conv}e, we present a circuit diagram of the Convolutional QFT. The key observation is that the {\em reversed} initial chain of $CX$ can be compiled significantly later in the circuit, compared to Fig \ref{fig:conv}a, and these $CX$ are no longer uniformly upstream of all layers of QFT rotation operations. After compiling the initial chain of $CX$ `as-late-as-possible' (ALAP) the circuit takes the form of a convolutional kernel (green dashed box). The translation-invariant gadget spans $d+2$ qubits. When sequentially incremented along the register, with a stride length of 1, the resulting unitary logic is the AQFT. The Fourier transform is calculated for the full register of length $n$ and emerges from the full convolution; the kernel gadget does {\em not} itself perform a smaller QFT in any individual stride or on any subset of qubits.

In this compilation strategy, qubits participate briefly in a dense kernel gadget. Otherwise, they are idle for long durations and protected from errors by dynamical decoupling. The decoupling sequences are not interrupted by additional algorithmic logic, since the logical operations are confined to the kernel. The Convolutional QFT is also less susceptible to gate-level noise and errors, since the average number of $CX$ gates in the causal history of each qubit wire is reduced. For example, the topmost qubit wire in Fig \ref{fig:conv}e is affected by only nine upstream $CX$ gates, reduced from 14 in Figs \ref{fig:conv}a, \ref{fig:simplified}a, and \ref{fig:simplified}b. Measurement operations with fewer causally upstream $CX$ gates are more likely to return the ideal result, since each gate contributes a certain amount of noise on average.

\section{Experimental Demonstration} \label{sec:experimental}

We validate the Convolutional QFT compilation strategy by executing QFT benchmarking circuits \cite{lubinski2023application} on the IBM Quantum Platform \cite{IBMQuantumPlatform}. For our experiments we select the Heron R3 device {\em ibm\_boston} due to low reported median two-qubit gate error. All data presented here use identical compilation and error suppression pipelines, shot counts, and back-to-back execution in two IBM Quantum Platform jobs to mitigate the impact of hardware drifts on the results we present. All target bitstrings are randomly selected.

We compile QFT circuits for different $n$ using the Convolutional QFT strategy and approximation via truncation of rotations of $\pi/16$ radians and below. This level of AQFT approximation reduces two-qubit gate counts dramatically at the expense of introducing synthesis error compared to the full QFT unitary definition in Eq \eqref{eq:qft_unitary}. The chosen threshold maximizes QFT process fidelity in the current benchmarking exercise on the selected device. 

Transpilation for the IBM Heron devices replaces each $CX$ with one $CZ$ plus additional single-qubit gates, preserving the two-qubit gate count of the input circuit. Gate scheduling is `as late as possible' (ALAP), with one key exception: the final entangling gate on each qubit pair is moved forward as far as possible, matching the schematic in Figure \ref{fig:conv}c. This scheduling policy ensures that each instance of the convolutional kernel is as localized in time as possible. Measurements are appended ALAP, and are synchronized at the end of the circuit. We choose a layout as a linear subgraph of the full device containing on average the highest fidelity rotation and entangling gates \cite{mundada2023experimental,wang2025delta}. Finally, dynamical decoupling is inserted into the idle delays using a robust crosstalk-suppressing embedding strategy \cite{coote2025resource}.  All circuits are submitted to the IBM Quantum Platform using identical runtime settings.

\begin{figure*}
    \centering
    \includegraphics[width=0.99\linewidth]{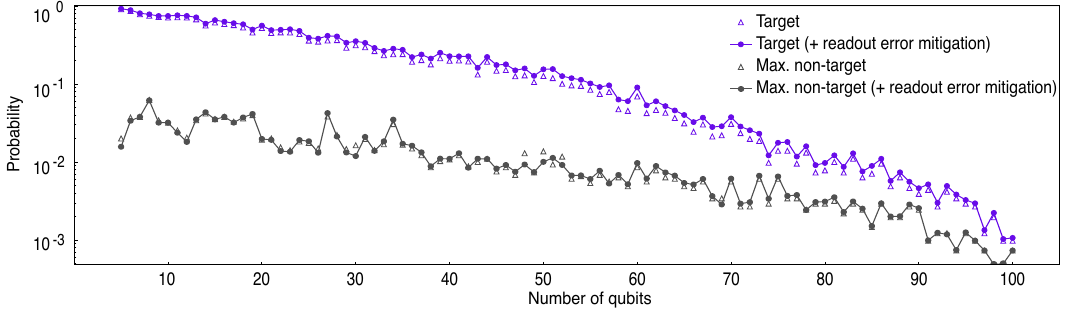}
    \caption{Convolutional QFT circuit success probability for circuit widths of 5 to 100 qubits. In each experiment, a periodic quantum state with a randomly selected frequency is passed into the convolutional QFT circuit, and the output is measured. For all circuit sizes, the correct integer target frequency is the unique highest likelihood result in both the raw data and after the application of readout error mitigation (purple curve). The most prominent {\em incorrect} result for each circuit execution (gray curve) had lower observed likelihood compared to the correct target bitstring. Data were acquired on {ibm\_boston} with 4096 shots per circuit.
    }
    \label{fig:sweep100qubits}
\end{figure*}

To evaluate QFT circuit performance, we encode a periodic signal in a quantum state in $2^n$ probability amplitudes of an $n$-qubit register. This periodic state preparation requires only single-qubit gates: a layer of $H$ gates followed by $RZ$ rotations parameterized by the chosen target frequency \cite{nielsen2010quantum,lubinski2023application}. The target frequencies are sampled uniformly at random in all cases. For each trial circuit, the shallow state preparation layer is prepended to the compiled QFT, and no other modifications are made to the circuit, layout, dynamical decoupling sequences, or scheduling. Therefore, the unitary QFT sub-circuit is identical for all test circuits of the same width, and we test its operation on different elements randomly sampled from a sinusoidal basis. 

The ideal output is a unique element of the computational basis, which is then ideally measured as a specific `target' bitstring corresponding exactly to the randomly selected integer frequency. In principle, all such basis elements could be tested. In practice, it is only feasible to experimentally execute a small sample set, from which we can statistically estimate the process fidelity\cite{baumer2024quantum,aumann2026demonstrating}. Specifically, for each of $m=20$ randomly selected frequencies, the algorithm is executed with an observed success likelihood $p_k$ of returning the correct frequency. We then use the formula originally derived in \cite{baumer2024quantum} for an unbiased estimate of the process fidelity:
\begin{align}
    \hat{\mathcal{F}} = \frac{m}{m-1} \left( \frac{1}{m}\sum_{k=1}^{m} \sqrt{p_k}\right)^2 - \frac{1}{m(m-1)} \sum_{k=1}^{m} p_k \label{eq:fidelity}
\end{align}

In hardware execution we compare the raw measurement outcomes against the target bitstrings, evaluating the process fidelity of the composite `QFT + measurement' operation~\cite{baumer2024quantum}. To isolate the fidelity of the QFT unitary, we apply readout-error mitigation~\citep{cai2023quantum,maciejewski2020mitigation,mundada2023experimental}, which is particularly effective as misclassification errors compound across large qubit registers. 

We assess algorithmic performance as a function of register width $n$ (Fig \ref{fig:sweep100qubits}), compiling QFT circuits across a range of register sizes from $n = 5$ to $n = 100$ qubits, and prepending a state-preparation layer corresponding to a target frequency sampled uniformly at random from $[0, 2^n)$.   We present the probability of returning the target bitstring as well as the probability of the next-most-frequently returned bitstring.  In all cases to $n=100$ qubits, the mode measurement outcome consistently identifies the correct integer frequency of the input periodic state, and is distinguishable from any other candidate, with signal-to-noise greater than unity in every experimental dataset.  Here, the success probability remains above the previously articulated $\sim1\%$ threshold up to 80 qubits.

\begin{figure*}
    \centering
    \includegraphics[width=0.99\linewidth]{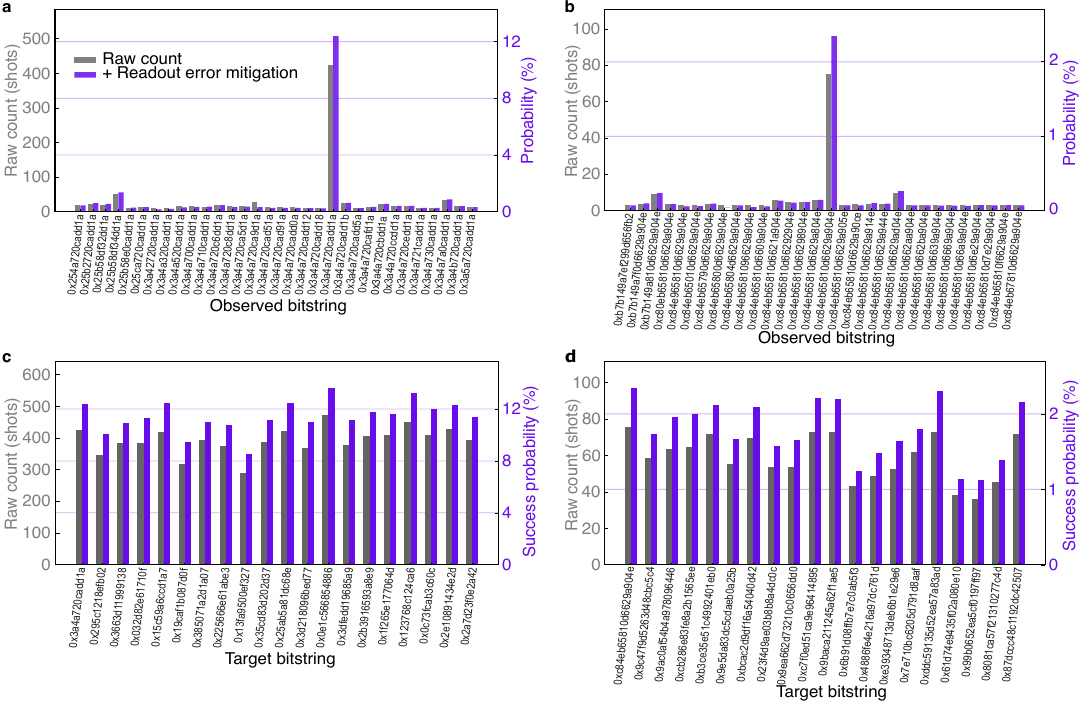}
    \caption{Experimental demonstration of the Convolutional QFT for 50 and 80 qubits. {\bf a., b.} Raw data, and data after mitigation of readout misclassification error, for a circuit width of 50 (a) and 80 (b) qubits. At both circuit widths, the mode observed bitstring is the correct ideal result. For visual clarity, only bitstrings with raw count of at least 10 shots (a) and 3 shots (b) are included. Bitstring labels are expressed in hexadecimal format due to the large number of bits. 
    {\bf c., d.}  Success probability (raw and readout-mitigated) for QFT circuits of size 50 (c) and 80 (d) qubits, each applied to 20 periodic quantum states with randomly selected frequencies. For all test circuits, the correct target is returned as the mode bitstring. Results from a, b are included in c,d as the leftmost columns, respectively. Circuit widths refer to the quantum register inputted to the QFT; we do not count the ancilla qubit as part of the `size' of the QFT demonstration.  Data are acquired on {ibm\_boston} with 4096 shots per circuit. 
    }
    \label{fig:50_80_qubits}
\end{figure*}

We can investigate the performance of these QFT circuits by examining the returned distributions for $n=50$ and $n=80$.  For both circuit widths the output distributions exhibit a sharp, distinct peak at the target bitstring, correctly reflecting the frequency encoded in the input state (Fig \ref{fig:50_80_qubits}a, b). For $n=50$ ($n=80$) the target bitstring is
observed $8.4\times$ ($7.5\times$) more frequently than any single incorrect output in the raw unmitigated data. Readout mitigation modestly increases the prominence of the correct bitstring, illustrated by the difference between gray and purple bars. 

We repeat experiments for both values of $n$ over 20 different randomly sampled target bitstrings, each time using the same compiled unitary QFT. The individual success rates for all 20 circuit trials are shown in Fig \ref{fig:50_80_qubits}c, d. Using Eq \eqref{eq:fidelity} and the raw data we estimate that the fidelity of `QFT + measure' for $n=50$ ($n=80$) is 9.6\% (1.4\%). From the data after applying readout mitigation, we estimate the process fidelity of the unitary QFT alone is 11.4\% (1.8\%). Fig \ref{fig:50_80_qubits}a is representative: in every trial the target bitstring was similarly elevated above the next most frequently observed measurement outcome, with the gap closing as $n\to100$. 

Inspecting the hexadecimal labels of the high-count outcomes in Fig \ref{fig:50_80_qubits}a, b reveals that many lie within a small Hamming distance of the target state. We emphasize that no classical post-processing (such as plurality voting \cite{maksymov2023enhancing}) is applied to harvest information from these near-target states; reported process fidelities reflect the direct success probability of the exact target bitstring.

\section{Conclusion}
We have presented and experimentally validated the Convolutional QFT compilation strategy on circuits with width up to 100 qubits. The construction is highly economical with two-qubit gates, making it suitable for experimental demonstration with a large number of qubits either in the pre-fault-tolerant era or in combination with QEC strategies.  Moreover, compilation for an LNN architecture has the advantage of being trivially compatible with the native device connectivity of all available hardware; for example, heavy-hex or square-lattice near-term devices, or grids of adjacent code patches in a fault-tolerant architecture. Furthermore, apart from a brief participation in the kernel, qubits are generally left idle and undisturbed by algorithmic operations, and protected from decoherence and crosstalk errors by uninterrupted dynamical decoupling sequences \cite{coote2025resource}.

The convolutional QFT can accommodate any desired threshold of approximate synthesis via small angle removal. For NISQ devices, we have demonstrated how optimizing the tradeoff between synthesis approximation and physical gate errors facilitates a demonstration with a high number of qubits. We emphasize that approximate synthesis is not limited to NISQ demonstrations; for future fault-tolerant devices, QFT implementations are likely to include only a small number of the largest defined rotations \cite{fowler2004scalability,nam2013scaling,nam2013streamlining,coppersmith2002approximate}, such as the $d=6$ largest angles \cite{white2023truncated}. For both current and future devices, the regime $d \ll n$ is highly relevant for practical compilation and resource estimation. In contrast, compilation strategies that accommodate only the {\em full} QFT \cite{klaver2026swap} ($d=n-1$) retain quadratic gate complexity despite the impracticality of realizing the required small rotations.

Using the convolutional strategy, we validated that the QFT subroutine executed on the IBM Quantum Platform can correctly identify the target state up to 100 qubits, with peak probability higher than any alternative incorrect bitstring. Consequently, given adequate shot statistics, the QFT successfully resolves the exact frequency of a periodic quantum state on registers up to and including 100 qubits.  This represents the largest QFT circuit successfully executed and capable of returning a computationally meaningful result. Furthermore, this QFT operates on a $2^{100}$-element signal-space dimension that far exceeds any classical Fourier transform implementation.

\subsection*{Acknowledgments}
We acknowledge the use of IBM Quantum Credits via the IBM Quantum Startups Program for this work. The views expressed are those of the authors and do not reflect the official policy or position of IBM or the IBM Quantum Platform team.  We are grateful to all our colleagues at Q-CTRL whose technical work has supported the results presented in this paper.

\subsection*{Author contribution statement}

PC designed and conducted the study. YB supervised the project. PC wrote the draft, and all authors reviewed and edited the manuscript. No AI or Large Language Models
were used in producing this work.

\subsection*{Data availability} 
All data are available upon reasonable request.

\bibliography{references.bib}

\end{document}